\documentclass{article}
\usepackage{arxiv}

\usepackage{microtype}
\usepackage{graphicx}
\usepackage{subfigure}
\usepackage{booktabs} 
\usepackage{algorithm}
\usepackage{algorithmic}
\usepackage{amsmath}
\usepackage{amssymb}
\usepackage{array}
\usepackage{threeparttable}
\usepackage{xcolor} 
\usepackage{booktabs}
\usepackage{graphicx}
\usepackage{multirow}
\usepackage{mathtools}
\usepackage{amsthm}
\usepackage[english]{babel}
\usepackage{pifont} 
\usepackage{multicol} 

\usepackage{hyperref}
\usepackage[utf8]{inputenc}
\usepackage{natbib}
\usepackage{pifont} 
\usepackage{caption} 

\theoremstyle{definition}

\theoremstyle{remark}

\usepackage[textsize=tiny]{todonotes}
\title{Invariant-based Robust Weights Watermark for Large Language Models }
\date{}

\author{
Qingxiao Guo,  Xinjie Zhu,  Yilong Ma,  Hui Jin,  Yunhao Wang,  Weifeng Zhang,  Xiaobing Guo \\
Lenovo Research, Beijing, China \\
Email: \{guoqx2, zhuxj11, mayl9, jinhui8, wangyh43, weifengz, guoxba\}@lenovo.com
}

\renewcommand{\headeright}{}
\renewcommand{\undertitle}{}

\hypersetup{
pdftitle={Invariant-based Robust Weights Watermark for Large Language Models},
pdfsubject={cs.CR, cs.LG},
pdfauthor={Qingxiao Guo, Xinjie Zhu, Yilong Ma, Hui Jin, Yunhao Wang, Weifeng Zhang, Xiaobing Guo},
pdfkeywords={Large language model, Watermarking, Copyright protection},
}

\begin{document}
\maketitle

\begin{abstract}
Watermarking technology has gained significant attention due to the increasing importance of intellectual property (IP) rights, particularly with the growing deployment of large language models (LLMs) on billions resource-constrained edge devices. To counter the potential threats of IP theft by malicious users, this paper introduces a robust watermarking scheme without retraining or fine-tuning for transformer models. The scheme generates a unique key for each user and derives a stable watermark value by solving linear constraints constructed from model invariants. Moreover, this technology utilizes noise mechanism to hide watermark locations in multi-user scenarios against collusion attack. This paper evaluates the approach on three popular models (Llama3, Phi3, Gemma), and the experimental results confirm the strong robustness across a range of attack methods (fine-tuning, pruning, quantization, permutation, scaling, reversible matrix and collusion attacks).
\end{abstract}

\keywords{Large language model, Watermarking, Copyright protection}

\section{INTRODUCTION}
\label{submission}

In recent years, large language models (LLMs), such as BERT \cite{kenton2019bert}, the GPT series \cite{achiam2023gpt} and LLaMA \cite{touvron2023llama}, have emerged as key technologies driving advancements in Natural Language Processing (NLP) \cite{nadkarni2011natural}. These Transformer-based models are applied across multiple domains \cite{costa2022no,hendy2023good,ni2023lever,vaithilingam2022expectation}. However, the unregulated distribution of LLMs has given rise to significant concerns regarding model misuse, unauthorized reproduction, and the potential for malicious redistribution, all of which threaten intellectual property security and accountability. \cite{chen2023can,perez2202red}.

To address these concerns, watermarking techniques for large language models (LLMs)  have been proposed as a means of protecting the intellectual property of LLMs. Watermarking techniques can be broadly classified into output-based and parameter-based schemes, depending on the location of the embedded watermark. Output-based watermarking schemes embed subtle patterns within the text generated by the model, enabling content traceability even without direct access to the model \cite{kirchenbauer2023watermark, liu2024adaptive}. However, these methods are primarily designed to track the provenance of the generated output and do not address the protection of the intellectual property (IP) of the model parameters. Moreover, output-based approaches are vulnerable in on-device scenarios, where adversaries can detect, manipulate, or remove watermarks through local analysis. In contrast, parameter-based watermarking schemes aim to safeguard model ownership by embedding watermarks directly into the model's parameters \cite{pang2024adaptive, gu2022watermarking}. These approaches provide inference-invisible solutions that are inherently resistant to detection and tampering at the output level.

This paper introduces an innovative robust watermarking scheme, categorized as a parameter-based watermark, specifically tailored for on-device deployments. Since the watermark is embedded directly into the model's weights, this approach is referred to as weights watermark. It offers a foundational solution to protect the intellectual property (IP) of the models without compromising their performance or usability. Weights watermarking generates a unique embedded key for each user, and obtains a relatively stable watermark value from the model invariants, ensuring the uniqueness of the watermark, providing a strong tracking ability to trace back to specific suspicious users, and enhancing IP protection in multi-user scenarios. Moreover, the scheme enhances the robustness to collusive attacks by introducing noise in multi-user scenarios to mask the watermark location and increase the resilience to collusive attacks. 

In summary, our contributions are summarized as follows:
\begin{itemize}
    \item We introduce a novel weights watermarking scheme. By computing the condition number of the model invariant, a stable watermark value is obtained, which enables a robust watermarking framework without requiring retraining or fine-tuning for transformer models.
    \item Our proposed scheme uniquely enables the retrieval of watermark values. A permutation matrix key is assigned to each user and the watermark value is derived by solving a linear system based on model invariants, linking it to the model parameters.
    \item We define and construct an anti-collusion multi-user watermarking scheme. It hides the positions of the watermarks among users by adding random noise.
    \item This paper evaluates the
approach on three popular models, and the experimental results confirm the strong robustness across a range of attack methods.
\end{itemize}

\section{RELATED WORK}

Prior arts weights watermark research focuses on DNNs with publicly accessible components during the model retraining or fine-tuning stage \cite{chen2020specmark}. Uchida \cite{uchida2017embedding} is the first to embed watermark information in the weight distribution of selected intermediate layers, but this method is vulnerable to rewriting attacks. To enhance security and robustness, DeepSigns \cite{darvish2019deepsigns} inserts watermarks into dynamic activation distributions, while DeepMarks \cite{chen2019deepmarks}  provides user vector encoding through weight regularization and anti-collusion codes. Other methods, such as those of \cite{kuribayashi2019efficient}, improve the confidentiality of the watermark using quantized index modulation, and \cite{feng2020watermarking} modulates the watermarks using orthogonal transformations. EmMark \cite{zhang2024emmark} proposes a robust watermarking framework for quantized embedded LLMs on edge devices. However, none of these methods effectively withstand the attacks described by \cite{yan2023rethinking} and \cite{pan2023cracking}. Although Fernandez's method maintains the watermark after specific modifications, but identity information can be easily erased \cite{fernandez2024}. 

Research on weights watermark for large language models is limited. Most of these methods involve retraining or fine-tuning, leading to high computational costs. These watermarking schemes exhibit insufficient robustness and struggle to counter various attacks, including fine-tuning, pruning, quantization, permutation, scaling, inverse transformations, and collusion. Therefore, we present a robust weights watermark scheme as the first watermarking framework without retraining and fine-tuning for
embedded LLM and protecting owners’ IP in the edge.

\section{THREAD MODEL AND ATTACK}
Effective watermarking technology should have the following key characteristics: fidelity, robustness, efficiency, detectability, and traceability \cite{chen2019deepmarks}. Consistent with the assumptions made by \cite{feng2024aqualora} and \cite{zhang2024emmark}, we assume that attackers have full access to the watermarked embedded LLMs parameters and possess some knowledge of the watermark embedding algorithm. However, attackers cannot obtain the original weights and are unaware of the random keys used by the owner to select watermark parameters.

Currently, the main attack strategies against watermarking frameworks include ambiguity attacks and removal attacks \cite{TAN2024}. An ambiguity attack (also known as a confusion attack) refers to the embedding of another illegal watermark in the model, which compromises the uniqueness of the original watermark and confuses its authenticity \cite{wu2004collusion}. However, removal attacks aim to erase the watermark information from the protected model, rendering the verification process invalid. Although operations such as fine-tuning, pruning, or quantization can inadvertently remove the watermark, malicious removal attacks can be implemented through transformation attacks \cite{adi2018turning,lukas2022sok}. Transformation attacks involve rearranging, scaling, or applying reversible transformations to model parameters to disrupt the embedded watermark signal. Common forms of transformation attacks include permutation attacks, scaling attacks, and reversible matrix attacks. Since these attacks are closely related to the model's structure, a detailed discussion is provided in Appendix \ref{appendixA} and \ref{appendixB}.

Collusion attacks are common techniques in watermark removal attacks. If the design of the watermark is flawed, some users may succeed in conspiring to generate a set of model parameters devoid of detectable watermark traces. Therefore, there is a pressing need for a watermarking scheme that can identify attackers and resist collusion attacks to protect model content. Common forms of collusion attacks include linear collusion attacks, where users synchronize their watermarked copies and take the average, and copy-paste attacks, where users cut and paste parts of each other’s models to form a new model \cite{ergun1999note}. Other attack methods may involve non-linear operations, such as taking the maximum or median value of the corresponding components from individual copies \cite{kilian1998resistance,cox1996secure}.

\begin{table}
\centering
\caption{Robustness comparison of watermarking methods under different attacks (Sin.: Single-user, Mul.: Multi-user, FT: Fine-Tuning, Prun.: Pruning, Quan.: Quantization, Perm.: Permutation, Scal.: Scaling, Revers.: Reversible, Col.: Collusion).}
\resizebox{14cm}{!}{ 
\begin{tabular}{l@{\hspace{3pt}} c@{\hspace{3pt}} c@{\hspace{3pt}} c@{\hspace{3pt}} c@{\hspace{3pt}} c@{\hspace{3pt}} c@{\hspace{3pt}} c@{\hspace{3pt}} c@{\hspace{3pt}} c@{\hspace{3pt}}}
    \toprule
    \textbf{Models} & \textbf{Sin.} & \textbf{Mul.} & \textbf{FT} & \textbf{Prun.} & \textbf{Quan.} & \textbf{Perm.} & \textbf{Scal.} & \textbf{Revers.} & \textbf{Col.} \\ 
    \midrule
    {EmMark \cite{zhang2024emmark}} & \ding{51} & \ding{51} & \ding{51} & \ding{55} & \ding{51} & \ding{55} & \ding{55} & \ding{55} & \ding{55}\\  
    {Fernandez \cite{fernandez2024}}  & \ding{51} & \ding{51} & \ding{51} & \ding{51} & \ding{51} & \ding{55} & \ding{51} & \ding{51} & \ding{55}\\ 
    {Ours}   & \ding{51} & \ding{51} & \ding{51} & \ding{51} & \ding{51} & \ding{51} & \ding{51} & \ding{51} & \ding{51}\\ 
    \bottomrule
\end{tabular}
}
\label{tab:attack-methods}
\begin{tablenotes}
\footnotesize
\item Note: \ding{51}: Resistant,  \ding{55}: Not Resistant
\end{tablenotes}
\end{table}

This solution successfully resists every attack, including fine-tuning, pruning, quantization, permutation, scaling, reversible, and collusion. It shows superior robustness in all types of attacks compared to previous methods, as shown in \hyperref[tab:attack-methods]{Table~\ref{tab:attack-methods}}. Section \ref{sec:5.3} provides detailed attack experiments to validate these conclusions.

\section{THE PROPOSED SCHEME}
\subsection{Overview}
\begin{figure}[h]
  \centering
  \includegraphics[width=\linewidth]{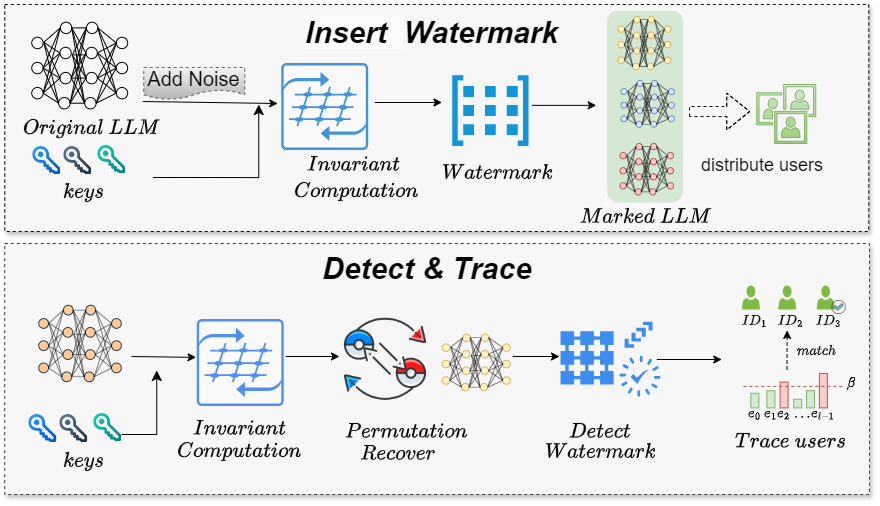}
  \caption{In the insertion stage, invariants are calculated, orthogonal noise is optionally added, and the watermark is embedded. In the detection stage, invariants are recalculated, the permuted weight matrix is restored, and watermark detection is performed.}
  \label{fig:ir_mark} 
\end{figure}

This solution consists of two main phases: watermark insertion and watermark detection, as shown in \hyperref[fig:ir_mark]{Figure~\ref*{fig:ir_mark}}. In the embedding phase, the model owner randomly selects the row vectors from the embedding layer weight matrix $W_e$ and calculates invariant information based on model weight data and private watermark key. Next, the model owner randomly selects different row vectors in $W_e$ as target vectors for the embedding of the watermark. The watermark values are determined by solving a system of linear equations, and after proper scaling, they are inserted into the weight matrix of the embedding layer. In the detection stage, the model owner extracts the watermark to trace the real owner of the model. By extracting the invariant matrix from the suspicious model and simultaneously identifying the vectors with embedded watermarks in the embedded layer, the final watermark score is computed. The model owner then compares and ranks these scores against random vectors to determine the final value of the watermark.

We provide single-user and multi-user modes depending on the application scenario. The single-user mode offers better performance and robustness, while the multi-user mode is more adaptable to a broader range of application scenarios.

\subsection{Single-User}

\begin{figure*}
  \centering
  \includegraphics[width=\linewidth]{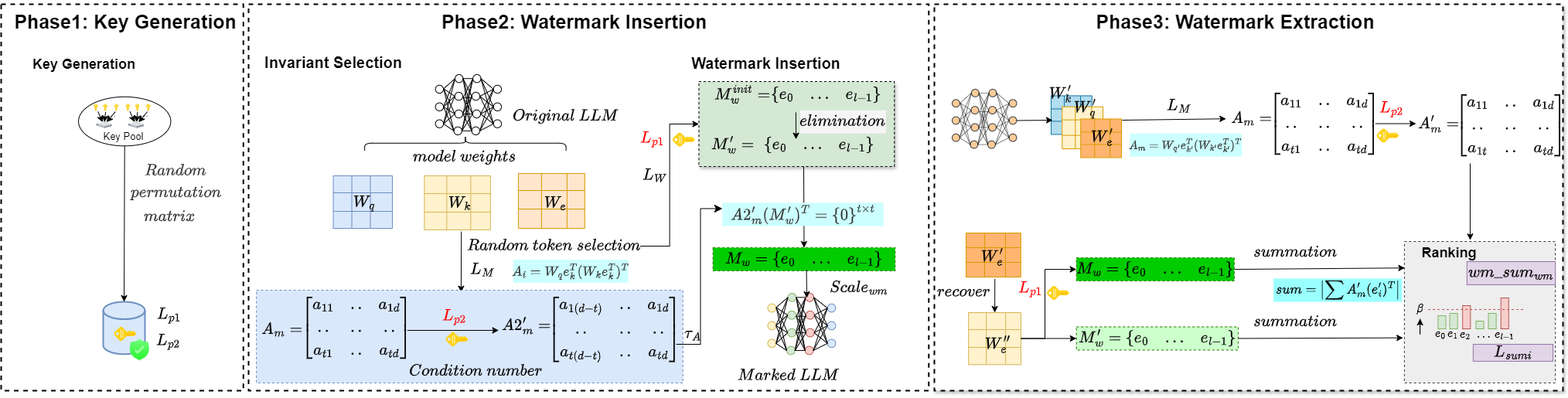}
  \caption{Inserting and extracting watermarks in single-user scenarios. In phase 1, keys can be generated offline and stored securely. In phase 2, linear equations are employed to obtain watermarks. In phase 3, the model owner extracts the watermark and traces the true owner of the model by matching the result with the locally stored private key.}
  \label{fig:algorithm} 
\end{figure*}

In the single-user scheme, the process is streamlined and foundational, focusing solely on extracting the watermark from a single source. As shown in \hyperref[fig:algorithm]{Figure~\ref*{fig:algorithm}}, it consists of three main phases: key generation, watermark insertion, and watermark detection. In the key generation phase, the model owner generates multiple permutation matrices as private keys for embedding the watermark, securely storing them locally for later extraction. In the watermark insertion phase, the embedding layer of the Transformer model is chosen for watermarking because embedding the watermark in the Attention or MLP layers could significantly degrade performance during inference. Additionally, embedding watermarks in this layer helps protect against functional equivalence attacks \cite{uchida2017embedding, liu2021watermarking, lukas2022sok}, such as those using invertible matrix transformations. In the watermark extraction phase, the model owner extracts the watermark by identifying the invariant matrix and the marked vectors in the embedding layer. The final watermark score is computed and compared against random vectors, with the results matched to the stored private keys for precise user attribution.

\subsubsection{Invariant Selection}
As shown by \cite{uchida2017embedding, liu2021watermarking, lukas2022sok} functionality-equality attack can eliminate such embedded watermark without significant performance loss or dramatic change to the protected model. We select \( QK^T \)  as the basis for the invariant in the watermarking scheme. The detailed method for calculating this invariant shown in \hyperref[alg:calculate_invariants]{Algorithm~\ref{alg:calculate_invariants}} can be found in Appendix \ref{appendixC}.

\subsubsection{Watermark Insertion} The process of watermark insertion is as follows:\

\paragraph{Select the watermark insertion position.}
We embed the watermark into the row vectors of the embedding weight matrix \(W_e\). Specifically, we select \(l\) positions corresponding to certain tokens to insert the watermark. To minimize the impact, we can choose token positions that are used less frequently. However, this could allow an attacker to remove the weights of these tokens by trimming the model, effectively erasing the watermark. Yet this action would greatly impair the model’s ability to process unfamiliar tokens. For better robustness, we choose to insert the watermark into randomly selected positions avoiding using token positions required for computing the invariant matrix \(A_m\) (\(L_M\)) to prevent the watermark from damaging the invariants. According to these rules, the list of watermark insertion positions can be written as:
\[
L_W = [\text{ind}_1, \dots, \text{ind}_l \mid \text{ind}_i \in [0, s-1]]
\]

\paragraph{Calculate the watermark value.}

In this scheme,  watermark value calculation is based on a linear constraint condition. Specifically, for the watermark matrix \(M_w \in \mathbb{R}^{t \times d}\), we aim for the matrix multiplication \(B = A_m M_w^T\) to result in a zero matrix, i.e., \(B \in \{0\}^{t \times t}\). As a result, the process of calculating the watermark value boils down to solving a system of linear equations. The matrix \(M_w\) is created by concatenating row vectors from \(W_e\) based on the selected watermark insertion positions. Each row vector is permuted using a random permutation matrix,  with certain positions left as unknowns to be solved. The permutation matrix list used in this process will be stored as the secret key for watermark insertion. 

Despite the linear constraint, we want the watermark values to closely resemble the original weights to mask the watermark’s exact location and minimize performance degradation. To this end, we adopt two strategies: first, the calculated watermark values are scaled by a watermark scaling factor \(\text{scale}_{wm}\); second, we apply a permutation matrix to transform \(A_m\) into \(A_m'\), ensuring stability in the result distribution and avoiding extreme values. We measure the stability of the transformed results using the condition number of \(A_m'\). To prevent attackers from trimming specific row vectors in $W_e$ and potentially losing invariants, we can select a commonly used set of tokens. The random numbers in the $L_M$ list should be selected from this token set.

\paragraph{Insert the Watermark.}
Once the watermark matrix \(M_w\) is calculated, we update each corresponding row of the embedding weight matrix \(W_e\) with the rows from \(M_w\), yielding the final matrix \(W_e'\) containing the embedded watermark.The detailed computation process is shown in \hyperref[alg:watermark_embedding_single_user]{Algorithm~\ref{alg:watermark_embedding_single_user}} (in Appendix \ref{appendixC}).

\subsubsection{Watermark Extraction}
\label{sec:4.2.3}

The extraction process hinges on a crucial concept: the constraints imposed during watermark computation should remain stable, even under various attacks. This means that the result of the computation $A_m (M_w)^T$ should approximate zero.  However, since attackers might use permutation matrices in functional equivalence attacks, disrupting the calculation between the watermark and invariant matrices, we must first restore the order of the embedding layer weights in the watermarked model before extracting the watermark.

Specifically, we determine the order by calculating the cosine similarity between the embedding layer weights of the watermarked model $W_e'$ and the original model embedding layer weights $W_e$. For each weight, we choose the one with the highest similarity, thereby eliminating the effect of functional equivalence attacks caused by permutation matrices. Previous research has shown that this recovery method is reliable \cite{trias2024find}. 

Following the recovery operations on the watermarked model, we then utilize several random permutation matrices to \(M_w\) to derive \(M_w'\) and compute the value of $A_m (M_w')^T$. We can get the distribution of computation results across the various watermark positions and evaluate the ranking of results obtained with the watermark key within this distribution. If the number of watermark positions satisfies the ranking threshold, we may conclude that the watermark extraction is success.

For the specific watermark extraction algorithm, refer to \hyperref[alg:watermark_extraction]{Algorithm~\ref{alg:watermark_extraction}} 
(in Appendix \ref{appendixC}).

\subsection{Multi-User}
\label{sec:4.3}

In the multi-user scheme, our goal is to extract the watermark and identify the specific user associated with it. However, in a multi-user scenario, the attack surface is broader compared to a single-user context, as multiple malicious users may collude to locate the watermark positions \cite{wang2019attacks}, and selectively remove them through cropping, adding noise, etc.

\begin{figure}[h]
  \centering
  \includegraphics[width=\linewidth]{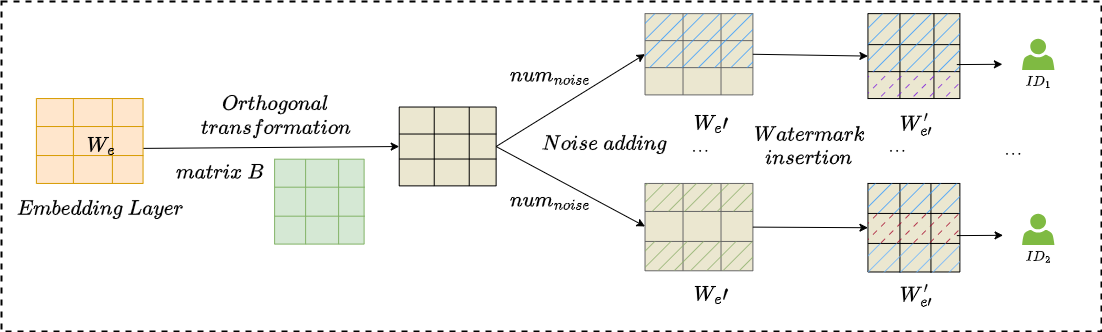}
  \caption{Multi-User noise adding and watermark insertion.}
  \label{fig:add_noise}
\end{figure}

To defend against such collusion attacks, as illustrated in \hyperref[fig:add_noise]{Figure~\ref*{fig:add_noise}}, we attempt to add the watermark more uniformly across its positions and introduce noise in non-watermark positions to obscure the exact watermark locations. This approach not only enhances the robustness of the watermark but also  significantly increases the difficulty for attackers in locating the watermark.

\subsubsection{Noise Adding}

In this scheme, we introduce noise at the non-watermark positions in the target model to mask the weight distribution differences introduced by the watermark. To minimize the noise's impact on model performance,  we aim to keep the second norm of the row vectors in the embedding weight matrix as stable as possible after adding the noise.  Given that the weight distributions across model layers typically follow a Gaussian distribution centered at 0, we assume the embedding layer's weights also follow a Gaussian distribution with parameters \((0, \sigma_E)\) \cite{lin2016fixed}.

We apply an orthogonal matrix to transform the model weights at non-watermark positions, setting certain weight values to \(\pm \sigma_E\). By performing the inverse orthogonal transformation, we can obtain the weights with added noise. Due to the property of orthogonal transformations that preserve the second norm, the second norm of the model weights remains approximately unchanged after the addition of noise. The specific operation flow is detailed in \hyperref[alg:noise_addition]{Algorithm~\ref{alg:noise_addition}} (in Appendix \ref{appendixC}).

\subsubsection{Invariant Selection}

The method for calculating invariants in the multi-user scheme is consistent with that in the single-user scheme. We use the noisy model embedding layer weights \(W_{e'}\) as input for Algorithm 1 to compute the invariant matrix \(A_m\).

\subsubsection{Watermark Insertion}

In the multi-user scheme, the watermark insertion process is similar to that of the single-user scheme, with the main difference being that the watermark matrix is calculated using weights after orthogonal transformation. The watermark insertion positions are selected randomly, just as in the single-user approach, to strengthen resilience against model pruning attacks.

Each user will receive a unique list of watermark insertion positions. The matrix list \(L_{(P_{u1})}\) used for watermark insertion will be stored as the key corresponding to user \(u\). After inserting the watermark in \(W_{e'}\), we can perform the inverse orthogonal transformation to get the watermark weights \(W_{e'}'\). The specific process is detailed in \hyperref[alg:insert_watermark_multiuser]{Algorithm~\ref{alg:insert_watermark_multiuser}} (in Appendix \ref{appendixC}).

\subsubsection{Watermark Extraction}

The multi-user watermark extraction scheme shown in \hyperref[alg:extract_watermark_multiuser]{Algorithm~\ref{alg:extract_watermark_multiuser}} (in Appendix \ref{appendixC}) resembles that of the single-user scheme. The key differences are: First, watermark extraction must be conducted on weights that have undergone orthogonal transformation. Second, the order of the computation results corresponding to their watermark positions must be obtained each user. If the ranking results for multiple users meet the threshold, the detected model is a combination of watermarked models, ensuring at least one user is identified. 

\section{EXPERIMENTS}

In this section, we evaluate the performance and robustness. First, we present a formula to calculate the watermark extraction success rate per watermark parameters, and then we select several target parameter sets. Next, we test the robustness of the single-user scheme and the multi-user scheme against several major attack methods. In the multi-user scheme, we evaluate the difficulty of identifying the watermark location under collusion attacks. Finally, we test the impact of the watermark on the model’s performance.  

\subsection{Experiment Setup}
We use the Gemma-2B, Phi3-4B, and Llama3-8B models for our experiments because their sizes make them suitable for edge-side applications. Additionally, their structural differences highlight the broad applicability of our method. The experiments were conducted on an NVIDIA A800 GPU with 80GB of RAM.

In our scheme, the adjustable parameters include the number of watermark locations \( l \), the watermark length at each location \( t \), the watermark scaling factor \( \text{scale}_{wm} \), the ranking threshold \( \beta \), the watermark detection threshold \( \rho \), and the number of users \( \text{num}_u \) in the multi-user scenario. During the experiments, we fixed certain parameters: \( l \) was set to 50, \( t \)  to \{5, 10, 15, 20\}, \( \beta \) to the range of 0.01 to 0.3 and  \( \rho \) to the range of 30 to 50. The parameters \( \text{scale}_{wm} \) and  \( \tau \) were adjusted dynamically based on model-specific weight distributions to effectively hide the watermark values.

Next, we present the formula for calculating the watermark extraction success rate. In the extraction phase,  \( p \) represents the ratio of locations that match the threshold. Using this ratio, along with the parameters \( l \) and \( \rho \), we calculate the success rate. Assuming each location meets the threshold independently, the success rate formula is:

\[
\text{Pr}_{WMsuccess} = 1 - \sum_{k=0}^{\rho - 1} \binom{l}{k} p^k (1 - p)^{l - k} \tag{1} \label{1}
\]
For the multi-user scheme, each user uses a different private key for each watermark location, we assume that when using a non-watermark user's private key for extraction, the probability of satisfying the ranking threshold is given by \( \beta \). Based on this, we first calculate the probability that any non-watermark user's private key satisfies the detection threshold \( \rho \), which is given by the formula:
\[
\text{Pr}_{Urandom} = 1 - \sum_{k=0}^{\rho - 1} \binom{l}{k} \beta^k (1 - \beta)^{l - k} \tag{2} \label{2}
\]
Accordingly, the probability that at least one non-watermark user's private key meets the detection threshold \( \rho \) when the number of users is \( \text{num}_u \) is:
\[
\text{Pr}_{Uwrong} = 1 - \left(1 - \text{Pr}_{Urandom}\right)^{\text{num}_u} \tag{3} \label{3}
\]
After providing the corresponding formulas for calculating the watermark extraction probability, we are ready to evaluate the performance and robustness in the next experimental phase.

\subsection{Performance}
\label{sec:5.2}
In this section, we will analyze the effectiveness, efficiency and fidelity.

\textbf{Effectiveness.} For effectiveness, we tested the watermark extraction success rates under two watermarking modes across different models, as shown in \hyperref[tab:detection]{Table~\ref*{tab:detection}}. The results indicate that a 100\% watermark extraction success rate was achieved in all modes for the three models tested.

\begin{table}[h]
    \centering
    \caption{Watermark detection rates for different models.}
    \footnotesize 
    \resizebox{8cm}{!}{
    \begin{tabular}{l c c | c c | c c} 
        \toprule
        \textbf{Model} & \multicolumn{2}{c|}{\textbf{Llama3-8B}} & \multicolumn{2}{c|}{\textbf{Phi3-4B}} & \multicolumn{2}{c}{\textbf{Gemma-2B}} \\
        \midrule
        Type & Sing. & Mul. & Sing. & Mul. & Sing. & Mul. \\
        \midrule
        Detect Rate & 100\% & 100\% & 100\% & 100\% & 100\% & 100\% \\
        \bottomrule
    \end{tabular}
    }
    \label{tab:detection}
\end{table}

\begin{figure}[h]
  \centering
  \includegraphics[width=\linewidth]{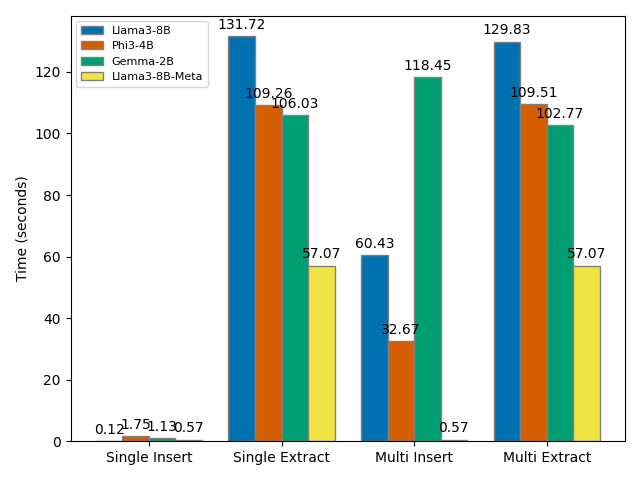}
  \caption{Watermark efficiency for insertion and extraction.}
  \label{fig:efficiency}
\end{figure}

\textbf{Efficiency.} To evaluate efficiency, we measured the running of the watermark embedding and extraction, as shown in \hyperref[fig:efficiency]{Figure~\ref*{fig:efficiency}}. The optimal hyperparameter settings were determined to be \( (t=10, wm\_num=50, scale\_wm=1000) \).
The results indicate that single-user schemes have the highest embedding efficiency, while multi-user schemes are slower due to noise addition, which can be optimized by pre-generating noise offline. Compared to \cite{fernandez2024}, our method excels in single insertion tasks on Llama3-8B. While the reference work shows superior efficiency in extraction and multi-insertion tasks, our approach balances speed and robustness across various scenarios. 

\textbf{Fidelity.} To evaluate fidelity, we compared the performance of the original version with that of the watermarking versions. Using OpenCompass \cite{contributors2023opencompass}. we evaluated performance across 10 distinct tasks, including CEval \cite{huang2024c}, MMLU \cite{hendrycks2020measuring}, WiC \cite{pilehvar2018wic}, WSC \cite{levesque2012winograd}, CoPA \cite{roemmele2011choice}, CB \cite{de2019commitmentbank}, BoolQ \cite{clark2019boolq}, PiQA \cite{bisk2020piqa}, MultiRC \cite{khashabi2018looking}, and HellaSwag \cite{zellers2019hellaswag}. These tasks were selected based on the criteria of comprehensiveness and popularity described in previous research \cite{wang2021gpt, gao2021framework, liu2023languages}. 

We report performance indicators at $t = 5$, $t = 10$, $t = 15$, and $t = 20$, where $t$ represents the watermark parameters. As \hyperref[tab:llama3]{Table~\ref*{tab:llama3}} shows, the results indicate negligible impact on model performance. More detailed test results (\hyperref[tab:phi3-4b]{Table~\ref*{tab:phi3-4b}} and \hyperref[tab:gemma-2b]{Table~\ref*{tab:gemma-2b}}) can be found in Appendix \ref{appendixD}.

\begin{table*}
    \centering
    \caption{Performance comparison of Llama3-8B across different benchmarks and configurations.}
    \begin{tabular}{c c c cccc cccc}
        \toprule
        \multirow{2}{*}{\textbf{Benchmark}} & \multirow{2}{*}{\textbf{Metric}} & \multirow{2}{*}{\textbf{Vanilla}} & \multicolumn{4}{c}{\textbf{Single}} & \multicolumn{4}{c}{\textbf{Multi}} \\
        \cmidrule(lr){4-7} \cmidrule(lr){8-11}
        & & & \textbf{t=5} & \textbf{t=10} & \textbf{t=15} & \textbf{t=20} & \textbf{t=5} & \textbf{t=10} & \textbf{t=15} & \textbf{t=20} \\
        \midrule
        ceval & 5-shot, top-1 & 52.23 & 52.23 & 52.23 & 52.23 & 52.23 & 52.71 & 51.9 & 51.8 & 52.41 \\
        mmlu & 5-shot, top-1 & 68.27 & 68.27 & 68.27 & 68.25 & 68.25 & 68.32 & 68.17 & 68.08 & 68.24 \\
        WiC & 0-shot & 36.83 & 37.15 & 37.15 & 37.15 & 37.15 & 39.5 & 38.09 & 37.3 & 38.24 \\
        WSC & 0-shot & 23.08 & 23.08 & 23.08 & 23.08 & 23.08 & 25.96 & 23.08 & 24.04 & 24.04 \\
        COPA & 0-shot & 93 & 93 & 93 & 93 & 93 & 94 & 94 & 96 & 94 \\
        CB & 0-shot & 62.5 & 62.5 & 62.5 & 62.5 & 62.5 & 60.71 & 60.71 & 62.5 & 58.93 \\
        BoolQ & 0-shot & 84.37 & 84.34 & 84.34 & 84.34 & 84.34 & 84.1 & 84.46 & 84.04 & 84.25 \\
        piqa & 0-shot & 78.4 & 78.35 & 78.35 & 78.35 & 78.35 & 78.07 & 78.4 & 78.24 & 78.67 \\
        MultiRC & 0-shot & 81.75 & 81.68 & 81.64 & 81.66 & 81.66 & 82.22 & 81.77 & 81.64 & 81.42 \\
        hellaswag & 0-shot & 75.17 & 75.21 & 75.2 & 75.2 & 75.2 & 75 & 75.35 & 75.23 & 75.35 \\
        \midrule
        \textbf{mean} & & 65.56 & 65.581 & 65.576 & 65.576 & 65.578 & 66.059 & 65.593 & 65.887 & 65.555 \\
        \bottomrule
    \end{tabular}
    \label{tab:llama3}
\end{table*}

\subsection{Robustness}
\label{sec:5.3}
In this section, we analyze the robustness of the watermarked models against pruning, quantization, and fine-tuning attacks. In the multi-user scheme, we evaluate the difficulty of identifying the watermark location under collusion attacks.

\begin{figure}[h]
  \centering
  \includegraphics[width=\linewidth]{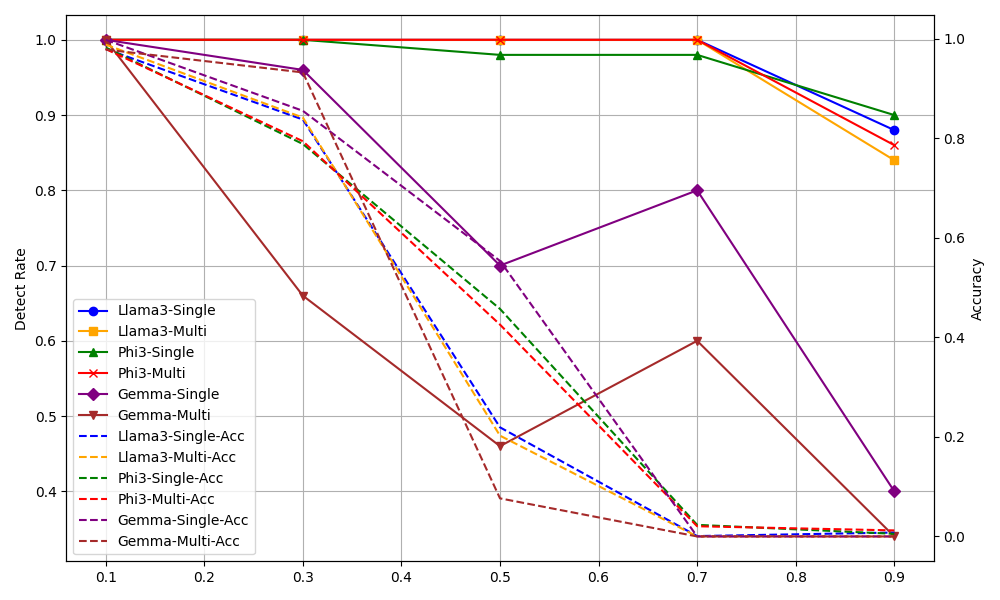}
  \caption{Detect rate and accuracy vs pruning rate.}
  \label{fig:prune}
\end{figure}

\textbf{Pruning.} As proposed in \cite{lukas2022sok}, pruning certain neurons can potentially remove the watermark from the model. We employed a global pruning method to evaluate the impact of pruning on watermark detection, where a pruning ratio \( r \) is set, and \( r\% \) of the neurons are zeroed out. The ratio \( r \) was varied from 0.1 to 0.9, with a step size of 0.2. We also measured the accuracy drop using CEval \cite{huang2024c} from OpenCompass \cite{contributors2023opencompass}, comparing different pruning levels to the original watermarked model. \hyperref[fig:prune]{Figure~\ref*{fig:prune}} shows that all models maintained nearly 100\% detection rates on a pruning ratio of 0.1. Beyond a 0.3 pruning ratio, accuracy drops significantly, rendering the detection rate declines irrelevant. These results demonstrate the robustness of the watermarking method against pruning attacks.

\textbf{Quantization.} In this section, we apply quantization techniques to single-user and multi-user schemes, both initially operating at 16-bit precision. We compress the models by converting their weights to 4-bit and 8-bit representations, which is considered a potential method for watermark removal \cite{lukas2022sok}. 

\begin{figure}[h]
  \centering
  \includegraphics[width=\linewidth]{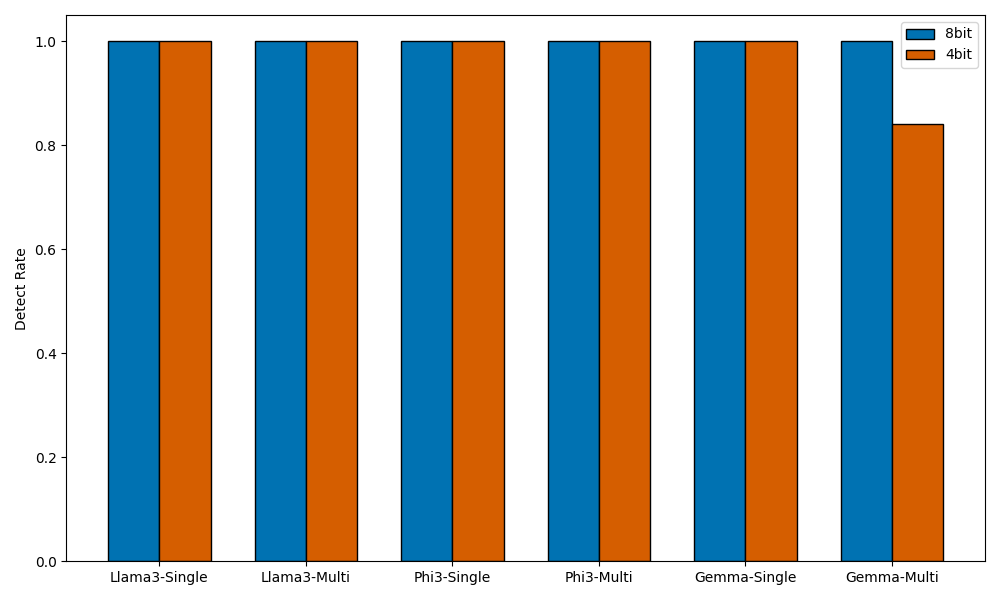}
  \caption{Detect rate vs quantize model.}
  \label{fig:quantize}
\end{figure}

As shown in \hyperref[fig:quantize]{Figure~\ref*{fig:quantize}}, our method maintains 100\% detection rates under all quantization levels, except for Gemma-2B at 4-bit in the multi-user case, where it drops to 0.84. Despite this reduction, detection remains effective, demonstrating our watermarking scheme’s resilience to quantization.

\textbf{Fine-tuning.} In the setup for fine-tuning attacks, we selected three datasets of varying sizes, including Blossom-Math-ZH \cite{blossom-math-v2}, Alpaca-ZH \cite{taori2023stanford}, and MS-Bench \cite{li2023modelscope}. For the MS-Bench dataset, we extracted 100k entries, allowing us to conduct fine-tuning attack tests using subsets of 10k, 50k, and 100k entries. The training was configured for 5 epochs with a learning rate of \(2 \times 10^{-5}\).

The results of the experiments are presented in \hyperref[tab:ft_results]{Table~\ref*{tab:ft_results}} (in Appendix \ref{appendixD}). We tested the distribution of watermark values under two different watermark lengths \(t = 10\), 20\, as well as the ranking threshold \(\beta\) at 1\%, 10\%, and 20\%. While the outcomes varied slightly among different models, the majority of watermark positions ranked within the top 20\%. Additionally, increasing the watermark length typically enhances the concentration of these values. Assuming an 80\% probability for ranks in the top 20\%, and based on equation~\eqref{1}, with a watermark count of 50, we set  \(\beta = 0.2\) \(\rho = 30\), resulting in a successful extraction rate of 99.96\%. Furthermore, according to equations~\eqref{2} and ~\eqref{3}, in a multi-user scenario with \(num_u = 10^5\), the probability of extracting a watermark from an incorrect user is 0.007\%. For models with relatively lower performance, such as Gemma-2B, the probability of ranking in the top 20\% is approximately 70\%. By adjusting \(\rho\) to 28, the successful extraction rate drops to 98.77\%, while the probability of extracting from an incorrect user at the same scale is 0.2\%.

\textbf{Collusion.} In our experiments, we simulated scenarios where multiple colluding malicious users attempted to remove the watermark by summing and averaging the weights, Afterward we used the keys of each malicious user to attempt watermark extraction from this model.

The experimental results are illustrated in \hyperref[fig:collusion]{Figure~\ref*{fig:collusion}} (in Appendix \ref{appendixD}). We varied the number of colluding users from 1 to 8 and compared the performance under different watermark lengths along with various scaling factors. More specifically, we tracked the probability of watermark positions ranking in the top 50\%. If this probability drops to 0.5, it indicates that the distribution of watermark positions is indistinguishable from a random distribution, meaning the watermark cannot be successfully extracted. 

From the results, it is evident that as the number of colluding users increases, the difficulty of watermark extraction also rises. For example, for the Llama3-8B model with \(t=5\) and \(scale\_wm=1000\), we could detect up to 8 malicious users. While the number of detectable malicious users fluctuated under different parameter settings, it is clear that the capacity for colluding users remains limited across various configurations. To maximize this threshold, we can adjust the watermark parameters based on the simulated results from different settings.

\textbf{Multiple Claims of Ownership.} In this ambiguity attack scenario, the adversary does not aim to remove the existing watermark, but rather to claim ownership of the model by embedding their own watermark. Given that the adversary only has access to the watermarked model \(M_1\), after generating their own watermark key, there are two primary attack strategies: (1) Embedding a watermark into the watermarked model \(M_1\), resulting in a doubly watermarked model \(M_{12}\), to claim ownership of the original model \(M_o\); (2) Performing a reverse computation on the watermarked model \(M_1\) based on their watermark values, yielding a new original model  \(M_{o}’\), then embedding the watermark into \(M_{o}’\), producing a watermarked model  \(M_{1}’\), to claim ownership of the model  \(M_{o}’\).

Under the first attack strategy, as the adversary lacks knowledge of the watermark embedding locations, they are relegated to random selection for watermark insertion. This has a negligible impact on the watermark within \(M_1\), which means the watermark of the owner can still be detected in the model \(M_{12}\). In the second attack strategy, although the adversary generates a new original model \(M_{o}’\), similar to the first scenario, the watermark of the owner remains detectable within \(M_{o}’\). In a situation of ownership conflict, the crux of the issue lies in which party can provide the unwatermarked original model and the watermarked model with a single watermark. Our proposed scheme effectively prevents the adversary from fulfilling these conditions, thereby resolving the ownership verification challenge.

\section{CONCLUSION}
We propose a lightweight robust watermarking scheme, which can provide traceability in multi-user scenarios for Transformer models. It binds the watermark to the model by identifying invariants within the Transformer architecture and extracting them as linear constraints on the watermark values. This mechanism ensures robustness against functionally equivalent attacks. Additionally, our scheme utilizes orthogonal transformations to introduce controlled noise, concealing the watermark values and providing resistance to collusion attacks within a limited group of attackers. The experimental results show that this scheme is robust against various known attacks, with low computational overhead, and has minimal impact on model performance. 

Future improvements will focus on expanding watermark insertion points by embedding watermarks in multiple layers, which will enhance resistance against pruning and fine-tuning attacks. Additionally, we plan to increase collusion tolerance by refining the noise mechanisms used in multi-user settings. Leveraging NFT technology will help ensure the uniqueness and traceability of the watermarks, further protecting against tampering and duplication attacks. These efforts will make the watermarking scheme even more robust and secure in real-world applications.

\section*{Impact Statement}
The protection of intellectual property (IP) in large language models (LLMs) is becoming increasingly critical as these models play an essential role in various applications, from natural language processing to complex decision-making systems. This paper addresses this challenge by proposing a novel watermarking mechanism to safeguard model parameters. Unlike traditional methods that focus solely on outputs, this paper proposes a novel method to protect the intellectual property (IP) of large language models (LLMs) by embedding secure, invisible watermarks into model parameters, ensuring ownership traceability, wide applicability across deployments, and minimal impact on performance.

\bibliographystyle{unsrtnat}
\bibliography{refs}

\newpage
\appendix
\onecolumn

\section{TRANSFORMER}
\label{appendixA}
The Transformer was introduced to address the limitations of RNNs and CNNs in handling sequence-to-sequence tasks. It relies entirely on the self-attention mechanism, allowing the model to process different parts of the input sequence in parallel, thereby improving computational efficiency and training speed. The structure of the Transformer typically includes the following modules: \(W_i\) represents the transformer weight matrix, \(\gamma_i\), \(\beta_i\) represent the transformer biases, and \(d\) is the dimension of the tokens.

\begin{enumerate}
    \item \textbf{Input Module}: Responsible for converting input data into a format that the model can process. This includes word embeddings (transforming words or subwords into fixed-dimensional vectors) and positional encodings (providing positional information of words in the sequence). This module includes the weight parameters \(W_e\) in the embedding layer.
     \[
    T_{\text{input}} = \text{Tokenization}(\text{input})
    \]
    \[
    x = T_{\text{input}} W_e, \quad W_e \in \mathbb{R}^{d \times s}
    \]

    \item \textbf{Attention Module}: First, the QKV projections are performed, with \(Q\), \(K\), and \(V\) being input to the softmax function, followed by processing through LayerNorm. Thus, the attention layer contains weight parameters \(\{W_q, W_k, W_v, W_o\}\) and biases \(\{\gamma_1, \beta_1\}\).

    \item \textbf{Output Module}: This module converts the model's internal representations into the final output and contains the output layer weight parameter \(W_c\).
    \[
    \text{output} = \text{SoftMax}(y W_c), \quad W_c \in \mathbb{R}^{d \times s}
    \]

\end{enumerate}

Additionally, the Transformer model typically has a multi-layer structure, with each layer using different parameters for the attention and feed-forward modules.

\section{ INVARIANT OF TRANSFORMER}
\label{appendixB}

After combining the formulas in Appendix A, we can systematically analyze how to attack parameter vectors by directly permuting the weights. In this process, the model's architecture remains unchanged, and the output is not affected. These attacks primarily take three forms: permutation attacks, scaling attacks, and invertible matrix attacks.

The key to a permutation attack lies in applying a permutation matrix. A permutation matrix is a special orthogonal matrix that has only one "1" per row and column, and its inverse is equal to its transpose, i.e., \(\pi^{-1} = \pi^T\). Among the weight parameters \(\{W_q, W_k, W_v, W_o, W_1, W_2\}\), there are three sets of permutation invariants: \(\{W_q, W_k\}\), \(\{W_v, W_o\}\), and \(\{W_1, W_2\}\).

To perform a permutation attack, we first apply the permutation operation on the word embedding matrix \(W_e\). Let \(X\) be the input, and we obtain:

\[
W_e' = W_e \pi, \quad X' = X W_e' = X W_e \pi
\]

Next, the invariants \(\{W_q, W_k\}\), \(\{W_v, W_o\}\), and \(\{W_1, W_2\}\) undergo permutation, scaling, and invertible matrix transformations.

\paragraph{Parameter Generation Process:}  
In the Transformer structure, this solution considers permutation, scaling, and invertible transformations, covering the entire process of parameter generation and transformation:

\begin{itemize}
    \item \textbf{Permutation Matrix Generation:}  
    Randomly generate a permutation matrix \(\pi \in \{0,1\}^{d \times d}\), and create a list of \(N\) permutation matrices:
    
    \[
    L_P = [\pi_1, \dots, \pi_N \mid \pi_i \in \{0,1\}^{d \times d}]
    \]

    \item \textbf{Invertible Matrix Generation:}  
    Randomly generate two lists, each containing \(N\) invertible matrices:

    \[
    L_B = [B_1, \dots, B_N \mid B_i \in \mathbb{R}^{d \times d}]
    \]
    \[\quad L_C = [C_1, \dots, C_N \mid C_i \in \mathbb{R}^{d \times d}]
    \]

    Then compute their inverse matrix lists:

    \[
    L_{B_{\text{inv}}} = [B_1^{-1}, \dots, B_N^{-1} \mid B_i^{-1} \in \mathbb{R}^{d \times d}]
    \]
    \[\quad L_{C_{\text{inv}}} = [C_1^{-1}, \dots, C_N^{-1} \mid C_i^{-1} \in \mathbb{R}^{d \times d}]
    \]

    \item \textbf{Scaling Factor Generation:}  
    Generate three lists of random scaling factors:

    \[
    L_a = [a_1, \dots, a_{N+1} \mid a_i \in \mathbb{R}]
    \]
    \[\quad L_d = [d_1, \dots, d_N \mid d_i \in \mathbb{R}]
    \]
    \[\quad L_e = [e_1, \dots, e_N \mid e_i \in \mathbb{R}]
    \]
\end{itemize}

\paragraph{Parameter Transformation Process:}
\begin{itemize}
    \item \textbf{Input Module:} The output of Tokenization, \(a_1\), is used as the input for the embedding layer. The weight matrix of the embedding layer undergoes row-column permutation using the permutation matrix \(\pi\):

\[
W_e' = a_1 W_e \pi
\]

\item \textbf{Attention Module:} Taking the parameters of the \(i\)-th layer as an example, the attention layer parameters undergo permutation, scaling, and invertible matrix transformations:

\[
W_{qi}' = \frac{1}{a_i} \pi^T W_{qi} B_i
\]
\[
W_{ki}' = \frac{1}{a_i} \pi^T W_{ki} (B_i^{\text{inv}})^T
\]
\[
W_{vi}' = \frac{1}{a_i} \pi^T W_{vi} C_i
\]
\[
W_{oi}' = a_i C_i^{\text{inv}} W_{oi} \pi
\]
\[
\gamma_{1i}' = d_i \gamma_{1i} \pi, \quad \beta_{1i}' = d_i \beta_{1i} \pi
\]

\item \textbf{Feedforward Module:} Taking the parameters of the \(i\)-th layer as an example, to maintain the performance of the Transformer, the feedforward layer parameters also need to undergo the corresponding permutation and scaling transformations:

\[
W_{1i}' = e_i \pi^T W_{1i} \pi_i
\]

\[
W_{2i}' = \frac{1}{e_i} \pi_i^T W_{2i} \pi
\]

\[
\gamma_{2i}' = a_{i+1} \gamma_{2i} \pi, \quad \beta_{2i}' = a_{i+1} \beta_{2i} \pi
\]

\item \textbf{Output Module:} To counteract the transformations applied to the Transformer structure, the output layer needs to multiply by the corresponding scaling factor and apply the inverse permutation matrix transformation:
\[
W_c' = \frac{1}{a_{N+1}} \pi^T W_c
\]
\end{itemize}

\section{ALGORITHMIC IMPLEMENTATION}
\label{appendixC}
The remaining algorithms for section \ref{sec:4.2.3} and \ref{sec:4.3} are presented below.


\begin{algorithm}
\caption{Invariant Selection}
\label{alg:calculate_invariants}

\raggedright \textbf{Input:} Embedding layer weights $W_e \in \mathbb{R}^{s \times d}$, Attention weights of the first layer $W_{q1}, W_{k1} \in \mathbb{R}^{d \times d}$ \par

\raggedright \textbf{Output:} Invariant matrix $A_m \in \mathbb{R}^{t \times d}$

\begin{algorithmic}[1]
    \STATE Randomly generate a list of $t$ random numbers: $L_M = [m_1, \dots, m_t \mid m_i \in [0, s-1]]$
    \FOR{$m_i \in L_M$}
        \STATE $e_i \leftarrow W_e[m_i]$
        \STATE $A_i \leftarrow e_i W_{q1} (e_i W_{k1})^T$
        \STATE $a_i \leftarrow A_i[0]$
    \ENDFOR
    \STATE Concatenate $a_1, \dots, a_t$ to form $A_m \in \mathbb{R}^{t \times d}$: $A_m \leftarrow \text{concat}(a_1, \dots, a_t)$
    \STATE \textbf{return} $A_m$
\end{algorithmic}
\end{algorithm}


\begin{algorithm}
\caption{Watermark Insertion in Single-User Scheme}
\label{alg:watermark_embedding_single_user}

\raggedright
\textbf{Input:} $W_e \in \mathbb{R}^{s \times d}$, $A_m \in \mathbb{R}^{t \times d}$, $L_W$, $\text{scale}_{wm}$, Condition number threshold $\tau$

\raggedright
\textbf{Output:} Watermarked embedding layer weights $W_e' \in \mathbb{R}^{s \times d}$, Matrix list $L_P1$, $L_P2$

\begin{algorithmic}[1]
    \FOR{$i = 0$ to max\_times}
        \FOR{$j = 0$ to $t-1$}
            \STATE $\pi_{Aj} \leftarrow \{0,1\}^{d \times d}$
            \STATE $A_m'[j] \leftarrow A_m[j] \pi_{Aj}$
        \ENDFOR
        \STATE $A2_m' \leftarrow A_m'[:, d-t:d]$
        \STATE $\tau_A \leftarrow \text{Condition\_number}(A2_m')$
        \IF{$\tau_A < \tau$}
            \STATE \textbf{break}
        \ENDIF
    \ENDFOR

    \STATE $L_P2 \leftarrow [\pi_{A0}, \dots, \pi_{A(t-1)}]$
    \STATE $L_P1 \leftarrow [\pi_{w1}, \dots, \pi_{wl} \mid \pi_{wi} \in \{0,1\}^{d \times d}]$
    
    \FOR{$i = 0$ to $l-1$}
        \STATE $e_i \leftarrow W_e[L_W[i]] L_P1[i]$
    \ENDFOR
    \STATE $M_w^{\text{init}} \leftarrow \text{concat}(e_0, \dots, e_{l-1})$
    \STATE $M_w' \leftarrow M_w^{\text{init}}$
    \STATE $M_w \leftarrow \text{Solve}(A_m (M_w')^T = \mathbf{0}^{t \times t})$
    \STATE $M_w \leftarrow \text{concat}(M_w[:, 0:d-t], (M_w[:, d-t:d]) / \text{scale}_{wm})$
    
    \FOR{$i = 0$ to $l-1$}
        \STATE $W_e'[L_W[i]] \leftarrow M_w[i] (\pi_{w1})^T$
    \ENDFOR

    \STATE \textbf{return} $W_e'$, $L_P1$, $L_P2$
\end{algorithmic}
\end{algorithm}

\begin{algorithm}[H] 
\caption{Watermarking Extraction in Single-User Scheme}
\label{alg:watermark_extraction}

\raggedright
\textbf{Input:}  $W_e \in \mathbb{R}^{s \times d}$, $W_e' \in \mathbb{R}^{s \times d}$, A $W_{q1}', W_{k1}' \in \mathbb{R}^{d \times d}$, Lists $L_M$, $L_W$,  $scale_{\text{wm}}$, $L_P1$, $L_P2$,  $\beta$,  $\rho$,  $numit$

\raggedright
\textbf{Output:} Watermark extraction result

\begin{algorithmic}[1]
    \STATE \textcolor{gray}{// Compute invariants} 
    \FOR{$m_i \in L_M$}
        \STATE $e_i \leftarrow W_e'[m_i]$
        \STATE $A_i \leftarrow e_i W_{q1}' (e_i W_{k1}')^T$
        \STATE $a_i \leftarrow A_i[0]$
    \ENDFOR
    \STATE $A_m \in \mathbb{R}^{t \times d} \leftarrow \text{concat}(a_1, \dots, a_t)$

    \STATE \textcolor{gray}{// Apply transformations to $A_m$}
    \FOR{$j = 0$ to $t-1$}
        \STATE $A_m'[j] \leftarrow A_m[j] L_P2[j]$
    \ENDFOR

    \STATE \textcolor{gray}{// Recover the original order of $W_e'$}
    \STATE $W_e'' \leftarrow \text{Permutation\_Recover}(W_e', W_e)$

    \STATE \textcolor{gray}{// Sum results for each watermark position}
    \FOR{$i = 0$ to $l-1$}
        \STATE $e_i \leftarrow W_e''[L_W[i]] L_P1[i]$
        \STATE $e_i' \leftarrow \text{concat}(e_i[0:d-t], e_i[d-t:d] \cdot scale_{\text{wm}})$
        \STATE $wm\text{sum}_i \leftarrow \text{abs}(\text{sum}(A_m' (e_i')^T))$
    \ENDFOR

    \STATE \textcolor{gray}{//Compute the distribution of products at each watermark position}
    \FOR{$i = 0$ to $l-1$}
        \FOR{$j = 0$ to $numit-1$}
            \STATE Randomly generate a permutation matrix: $\pi_j \in \{0,1\}^{d \times d}$
            \STATE $e_i \leftarrow W_e''[L_W[i]] \pi_j$
            \STATE $e_i' \leftarrow \text{concat}(e_i[0:d-t], e_i[d-t:d] \cdot scale_{\text{wm}})$
            \STATE $\text{sum}_j \leftarrow \text{abs}(\text{sum}(A_m' (e_i')^T))$
        \ENDFOR
        \STATE $L_{\text{sumi}} \leftarrow [\text{sum}_0, \dots, \text{sum}_{numit-1}]$
    \ENDFOR

    \STATE \textcolor{gray}{// Rank watermark positions}
    \FOR{$i = 0$ to $l-1$}
        \STATE $\text{order}_i \leftarrow \text{GetOrder}(wm\text{sum}_i, L_{\text{sumi}})$
    \ENDFOR

    \STATE \textcolor{gray}{// Count the number of watermark positions satisfying threshold conditions}
    \STATE $\text{detect} \leftarrow 0$
    \FOR{$i = 0$ to $l-1$}
        \IF{$\text{order}_i / numit < \beta$}
            \STATE $\text{detect} \leftarrow \text{detect} + 1$
        \ENDIF
    \ENDFOR

    \STATE \textcolor{gray}{// Watermark extraction success condition}
    \IF{$\text{detect} > \rho$}
        \STATE Watermark extraction is \textit{successful}
    \ELSE
        \STATE No watermark extracted from the target model
    \ENDIF
\end{algorithmic}
\end{algorithm}

\begin{algorithm}[H] 
\caption{Noise Addition}
\label{alg:noise_addition}

\raggedright
\textbf{Input:}  $W_e \in \mathbb{R}^{s \times d}$, $B \in \mathbb{R}^{d \times d}$, $num_{noise}$,  $L_W$ \par

\raggedright
\textbf{Output:} $W_{e'}$
\begin{algorithmic}[1]
    \STATE \textcolor{gray}{// Compute the standard deviation }
    \STATE $\sigma_E \leftarrow W_e$
    \STATE $\bar{W_e} \gets W_e B$
    \STATE \textcolor{gray}{// Add noise terms}
    \FOR{$i = 0$ to $s-1$}
        \IF{$i \notin L_W$}
            \STATE \textcolor{gray}{// Randomly generate $num_{noise}$ $[0, d-1]$}
            \FOR{$j = 0$ to $num_{noise}-1$}
                \IF{$j \% 2 == 0$}
                    \STATE $\bar{W_e}[i][nind_j] = \sigma_E$
                \ELSE
                    \STATE $\bar{W_e}[i][nind_j] = -\sigma_E$
                \ENDIF
            \ENDFOR
        \ENDIF
    \ENDFOR
    \STATE $W_{e'} \gets \bar{W_e} B^T$
\end{algorithmic}
\end{algorithm}

\begin{algorithm}[H]
\caption{Watermark Insertion in Multi-User Scheme}
\label{alg:insert_watermark_multiuser}

\raggedright
\textbf{Input:}  \(W_{(e')} \in \mathbb{R}^{s \times d}\),  \(A_m \in \mathbb{R}^{t \times d}\),  \(B \in \mathbb{R}^{d \times d}\),  \(L_W\), \(\text{scale}_{wm}\),  \(\tau\) \par

\raggedright
\textbf{Output:} \(W_{(e')}^{'} \in \mathbb{R}^{s \times d}\),  \(L_{(P_{u1})} = [\pi_{w1}, \ldots, \pi_{wl} \mid \pi_{wi} \in \{0,1\}^{d \times d}]\), \(L_P2 = [\pi_{A1}, \ldots, \pi_{At} \mid \pi_{Ai} \in \{0,1\}^{d \times d}]\)

\begin{algorithmic}[1]
    \STATE Steps 1-2 from Algorithm 2
    \STATE \textcolor{gray}{// Initialize watermark matrix}
    \FOR{i = 0 to l}
        \STATE \(e_i \leftarrow W_{(e')}[L_W[i]]B L_{P1}[i]\) 
        \STATE \(M_w^{init} \in \mathbb{R}^{t \times d} \leftarrow \text{concat}(e_0, \ldots, e_{l-1})\)
    \ENDFOR
    
    \STATE Step 4-6 from Algorithm 2
    
    \STATE \textcolor{gray}{//Overlay watermark matrix on original weights}
    \FOR{i = 0 to l}
        \STATE \(W_{(e')}^{'}[L_W[i]] \leftarrow M_w[i] L_{P1}[i]^T B^T\)
    \ENDFOR
    
    \STATE \textbf{return}  \(W_{(e')}^{'}, L_{(P_{u1})}, L_P2\)
\end{algorithmic}
\end{algorithm}

\begin{algorithm}[H]
\caption{Extract Watermark in Multi-User Scheme}
\label{alg:extract_watermark_multiuser}

\raggedright
\textbf{Input:} $W_e \in \mathbb{R}^{s \times d}$, $W_{e'}' \in \mathbb{R}^{s \times d}$, $W_{q1}', W_{k1}' \in \mathbb{R}^{d \times d}$, $B \in \mathbb{R}^{d \times d}$, $\theta_W = \{L_{W_{ui}} \mid i \in [0, \text{num}_u - 1] \}$, $\text{scale}_{wm}$, $\theta_{UK} = \{L_{P_{ui1}} \mid i \in [0, \text{num}_u - 1] \}$, $\theta_{AP} = \{L_{P_{ui2}} \mid i \in [0, \text{num}_u - 1] \}$, $\beta$, $\rho$, $\text{num}_u$, $\text{num}_{it}$ \par

\raggedright
\textbf{Output:} $\theta_{detect}$

\begin{algorithmic}[1]
    \STATE \textcolor{gray}{// Step 1: Transform } $A_m$
    \FOR{each $L_{P_{ux2}} \in \theta_{AP}$}
        \FOR{$j = 0 \ to \ t$}
            \STATE $A_{(m_{ux})}'[j] \gets A_m[j] L_{(P_{ux2})}[j]$
        \ENDFOR
    \ENDFOR
    
    \STATE \textcolor{gray}{// Recover the original permutation order of } $W_{(e')}'$
    \STATE $W_{(e')}' \gets \text{Permutation\_Recover}(W_{(e')}', W_e)$
    
    \STATE \textcolor{gray}{// Perform orthogonal transformation on } $W_{(e')}'$
    \STATE $\bar{W_{e}''} \gets W_{(e')}' B$
    
    \STATE \textcolor{gray}{// Calculate results at each watermark position using user keys and sum them}
    \FOR{each $L_{W_{ux}} \in \theta_W$ \textbf{and} $L_{P_{ux1}} \in \theta_{UK}$}
        \FOR{$i = 0 \ to \ l$}
            \STATE $e_i \gets \bar{W_{e}''}[L_{W_{ux}[i]}] L_{P_{ux1}}[i]$
            \STATE $e_i' \gets \text{concat}(e_i[0:d-t], e_i[d-t:d] \times \text{scale}_{wm})$
            \STATE $wm\sum_{(ux_i)} \gets \text{abs}(\text{sum}(A_{(m_{ux})}'(e_i')^T))$
        \ENDFOR
    \ENDFOR
    
    \STATE \textcolor{gray}{// Calculate distribution of the product of each user's watermark position with } $A_m'$
    \FOR{each $L_{W_{ux}} \in \theta_W$}
        \FOR{$i = 0 \ to \ l$}
            \FOR{$j = 0 \ to \ \text{num}_{it}$}
                \STATE $\pi_j \gets \text{Random}\{0, 1\}^{d \times d}$
                \STATE $e_i \gets \bar{W_{e}''}[L_{W_{ux}[i]}] \pi_j$
            \ENDFOR
            \STATE $e_i' \gets \text{concat}(e_i[0:d-t], e_i[d-t:d] \times \text{scale}_{wm})$
            \STATE $sum_j \gets \text{abs}(\text{sum}(A_{(m_{ux})}'(e_i')^T))$
            \STATE $L_{(sumi_{ux})} \gets [sum_0, \ldots, sum_{\text{num}_{it}-1}]$
        \ENDFOR
    \ENDFOR
    
    \STATE \textcolor{gray}{// Calculate ranking results for different users at each watermark position}
    \FOR{$i = 0 \ to \ l$}
        \STATE $order_{(ux_i)} \gets \text{GetOrder}(wm\sum_{(ux_i)}, L_{(sumi_{ux})})$
    \ENDFOR
    
    \STATE \textcolor{gray}{// Count the number of watermark positions meeting the threshold condition for different users}
    \STATE $detect_{ux} \gets 0$
    \FOR{$i = 0 \ to \ l$}
        \IF{$\frac{order_{(ux_i)}}{\text{num}_{it}} < \beta$}
            \STATE $detect_{ux} \gets detect_{ux} + 1$
        \ENDIF
    \ENDFOR
    
    \STATE \textbf{return} Set of users satisfying the condition $detect_{ux} > \rho$ as $\theta_{detect}$
\end{algorithmic}
\end{algorithm}

\section{EXPERIMENT RESULT}
\label{appendixD}
The results of experiment for sections \ref{sec:5.2} and \ref{sec:5.3} are presented below.


\begin{table*}[h]
    \centering
    \caption{Watermark detection results with fine-tuning (ranking threshold $\beta$ at 1\%, 10\% and 20\%).}
    \resizebox{12cm}{!}{ 
    \begin{tabular}{lccccccccc}
        \toprule
        \textbf{Model} & \textbf{t} & \textbf{Dataset size} & \multicolumn{3}{c}{\textbf{Single-User mode}} & \multicolumn{3}{c}{\textbf{Multi-User mode}} \\
        \cmidrule(lr){4-6} \cmidrule(lr){7-9}
         & & & 1\% & 10\% & 20\% & 1\% & 10\% & 20\% \\
        \midrule
        \multirow{6}{*}{Gemma-2B} 
         & \multirow{3}{*} & 10k  & 94\% & 96\% & 100\% & 100\% & 94\% & 100\% \\
         &                     & 50k  & 88\% & 92\% & 96\%  & 100\% & 88\% & 94\% \\
         &                     & 100k & 48\% & 74\% & 78\%  & 22\%  & 66\% & 76\% \\
         \cmidrule(lr){2-9}
         & \multirow{3}{*} & 10k  & 88\% & 100\% & 100\% & 60\%  & 64\% & 64\% \\
         &                     & 50k  & 60\% & 80\%  & 92\%  & 60\%  & 72\% & 76\% \\
         &                     & 100k & 40\% & 72\%  & 76\%  & 66\%  & 72\% & 82\% \\
        \midrule
        \multirow{6}{*}{Phi3-4B} 
         & \multirow{3}{*} & 10k  & 100\% & 100\% & 100\% & 100\% & 100\% & 100\% \\
         &                     & 50k  & 100\% & 100\% & 100\% & 100\% & 100\% & 100\% \\
         &                     & 100k & 100\% & 100\% & 100\% & 100\% & 100\% & 100\% \\
         \cmidrule(lr){2-9}
         & \multirow{3}{*} & 10k  & 100\% & 100\% & 100\% & 100\% & 100\% & 100\% \\
         &                     & 50k  & 100\% & 100\% & 100\% & 100\% & 100\% & 100\% \\
         &                     & 100k & 44\%  & 78\%  & 84\%  & 28\%  & 60\%  & 80\% \\
        \midrule
        \multirow{6}{*}{Llama3-8B} 
         & \multirow{3}{*} & 10k  & 100\% & 100\% & 100\% & 100\% & 100\% & 100\% \\
         &                     & 50k  & 100\% & 100\% & 100\% & 100\% & 100\% & 100\% \\
         &                     & 100k & 90\%  & 90\%  & 100\% & 88\%  & 94\%  & 100\% \\
         \cmidrule(lr){2-9}
         & \multirow{3}{*} & 10k  & 100\% & 100\% & 100\% & 100\% & 100\% & 100\% \\
         &                     & 50k  & 100\% & 100\% & 100\% & 100\% & 100\% & 100\% \\
         &                     & 100k & 58\%  & 58\%  & 76\%  & 68\%  & 78\%  & 82\% \\
        \bottomrule
    \end{tabular}
    }
    \label{tab:ft_results}
\end{table*}


\begin{table*}
    \centering
    \caption{Performance comparison of Phi3-4B across different benchmarks and configurations.}
    \resizebox{15cm}{!}{ 
    \begin{tabular}{c c c cccc cccc}
        \toprule
        \multirow{2}{*}{\textbf{Benchmark}} & \multirow{2}{*}{\textbf{Metric}} & \multirow{2}{*}{\textbf{Vanilla}} & \multicolumn{4}{c}{\textbf{Single}} & \multicolumn{4}{c}{\textbf{Multi}} \\
        \cmidrule(lr){4-7} \cmidrule(lr){8-11}
        & & & \textbf{t=5} & \textbf{t=10} & \textbf{t=15} & \textbf{t=20} & \textbf{t=5} & \textbf{t=10} & \textbf{t=15} & \textbf{t=20} \\
        \midrule
        ceval     & 5-shot, top-1 & 41.39 & 41.65 & 41.65 & 41.65 & 41.74 & 42.01 & 41.54 & 41.88 & 41.88 \\
        mmlu      & 5-shot, top-1 & 69.44 & 70.47 & 70.42 & 70.49 & 70.49 & 70.57 & 70.52 & 70.43 & 70.43 \\
        WiC       & 0-shot         & 60.66 & 65.20 & 65.20 & 65.20 & 65.20 & 65.20 & 65.05 & 65.52 & 65.52 \\
        WSC       & 0-shot         & 62.50 & 65.38 & 65.38 & 65.38 & 65.38 & 65.38 & 64.42 & 65.38 & 65.38 \\
        COPA      & 0-shot         & 97.00 & 98.00 & 98.00 & 98.00 & 98.00 & 97.00 & 98.00 & 98.00 & 98.00 \\
        CB        & 0-shot         & 66.07 & 71.43 & 71.43 & 71.43 & 71.43 & 67.86 & 71.43 & 71.43 & 73.21 \\
        BoolQ     & 0-shot         & 85.35 & 84.77 & 84.77 & 84.77 & 84.77 & 84.74 & 84.77 & 84.77 & 84.86 \\
        piqa      & 0-shot         & 79.00 & 83.46 & 83.51 & 83.46 & 83.51 & 83.35 & 83.35 & 83.19 & 83.19 \\
        MultiRC   & 0-shot         & 83.11 & 77.39 & 77.33 & 77.33 & 77.31 & 77.27 & 77.43 & 77.62 & 77.62 \\
        hellaswag  & 0-shot        & 77.79 & 79.41 & 79.42 & 79.42 & 79.42 & 79.24 & 79.31 & 79.39 & 79.39 \\
        \midrule
        \textbf{mean} & & 72.23 & 73.72 & 73.71 & 73.71 & 73.72 & 73.37 & 73.62 & 73.58 & 73.95 \\
        \bottomrule
    \end{tabular}
    } 
    \label{tab:phi3-4b}
\end{table*}

\begin{table*}
    \centering
    \caption{Performance comparison of Gemma-2B across different benchmarks and configurations.}
    \resizebox{15cm}{!}{ 
    \begin{tabular}{c c c cccc cccc}
        \toprule
        \multirow{2}{*}{\textbf{Benchmark}} & \multirow{2}{*}{\textbf{Metric}} & \multirow{2}{*}{\textbf{Vanilla}} & \multicolumn{4}{c}{\textbf{Single}} & \multicolumn{4}{c}{\textbf{Multi}} \\
        \cmidrule(lr){4-7} \cmidrule(lr){8-11}
        & & & \textbf{t=5} & \textbf{t=10} & \textbf{t=15} & \textbf{t=20} & \textbf{t=5} & \textbf{t=10} & \textbf{t=15} & \textbf{t=20} \\
        \midrule
        ceval     & 5-shot, top-1 & 31.97 & 31.97 & 31.97 & 31.97 & 31.97 & 33.98 & 30.17 & 30.79 & 30.79 \\
        mmlu      & 5-shot, top-1 & 35.25 & 35.25 & 35.25 & 35.24 & 35.24 & 34.76 & 33.79 & 34.99 & 34.99 \\
        WiC       & 0-shot         & 0.31  & 0.31  & 0.31  & 0.31  & 0.31  & 0.00  & 20.38 & 0.16  & 0.16  \\
        WSC       & 0-shot         & 0.96  & 0.96  & 0.96  & 0.96  & 0.96  & 0.00  & 28.85 & 0.96  & 0.96  \\
        COPA      & 0-shot         & 80.00 & 80.00 & 80.00 & 80.00 & 80.00 & 79.00 & 77.00 & 75.00 & 75.00 \\
        CB        & 0-shot         & 42.86 & 42.86 & 42.86 & 42.86 & 42.86 & 39.29 & 39.29 & 39.29 & 42.86 \\
        BoolQ     & 0-shot         & 4.95  & 4.95  & 4.95  & 4.95  & 4.95  & 0.21  & 10.52 & 5.75  & 5.75  \\
        piqa      & 0-shot         & 57.62 & 57.62 & 57.62 & 57.62 & 57.62 & 57.02 & 56.15 & 56.20 & 56.20 \\
        MultiRC   & 0-shot         & 38.72 & 38.72 & 38.72 & 38.72 & 38.72 & 38.59 & 38.30 & 37.97 & 37.97 \\
        hellaswag & 0-shot         & 36.67 & 36.67 & 36.67 & 36.67 & 36.67 & 35.98 & 33.88 & 35.71 & 35.71 \\
        \midrule
        \textbf{mean} & & 32.93 & 32.93 & 32.93 & 32.93 & 32.93 & 35.25 & 31.91 & 36.83 & 32.04 \\
        \bottomrule
    \end{tabular}
    } 
    \label{tab:gemma-2b}
\end{table*}

\begin{figure*}
    \centering
    \resizebox{15cm}{!}{
    \begin{minipage}{1.0\textwidth}
        \begin{minipage}{0.24\textwidth}
            \centering
            \includegraphics[width=\textwidth]{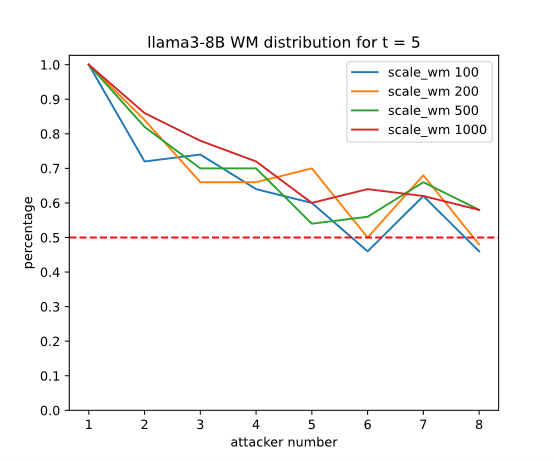}
        \end{minipage}
        \begin{minipage}{0.24\textwidth}
            \centering
            \includegraphics[width=\textwidth]{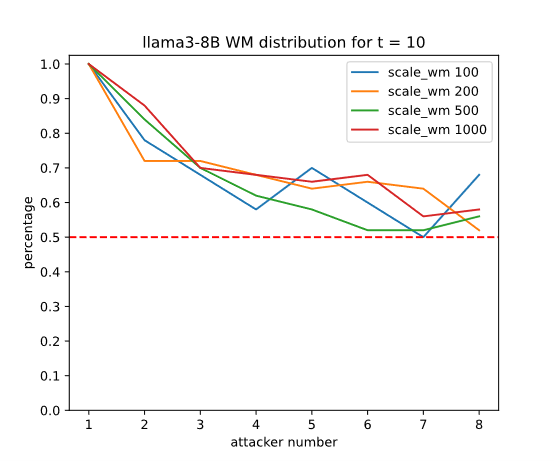}
        \end{minipage}
        \begin{minipage}{0.24\textwidth}
            \centering
            \includegraphics[width=\textwidth]{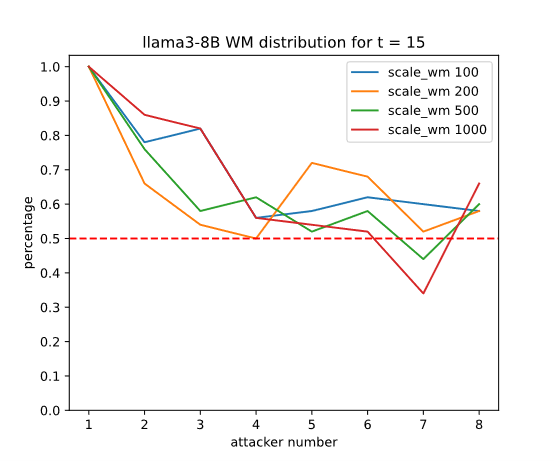}
        \end{minipage}
        \begin{minipage}{0.24\textwidth}
            \centering
            \includegraphics[width=\textwidth]{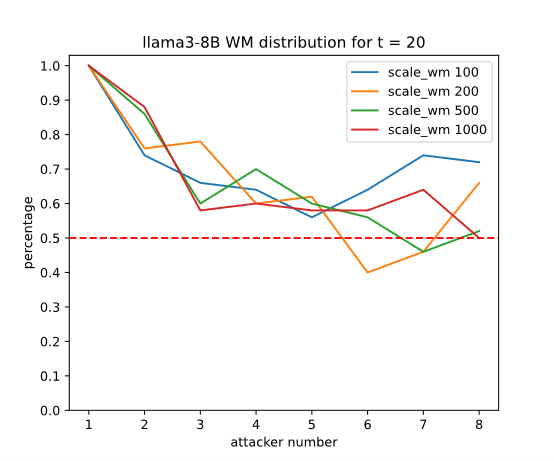}
        \end{minipage}
        
        \begin{minipage}{0.24\textwidth}
            \centering
            \includegraphics[width=\textwidth]{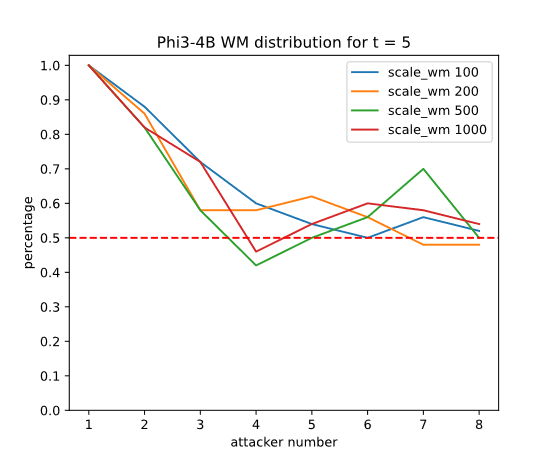}
        \end{minipage}
        \begin{minipage}{0.24\textwidth}
            \centering
            \includegraphics[width=\textwidth]{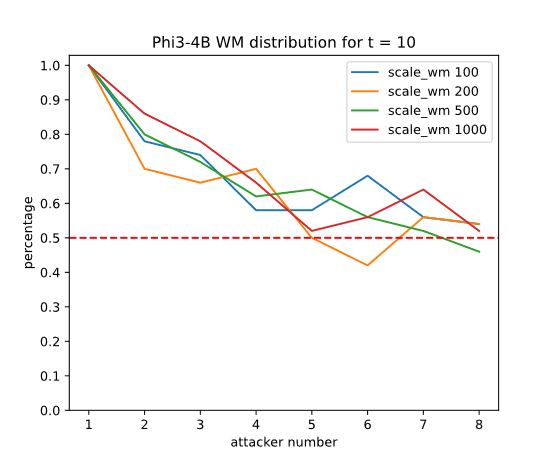}
        \end{minipage}
        \begin{minipage}{0.24\textwidth}
            \centering
            \includegraphics[width=\textwidth]{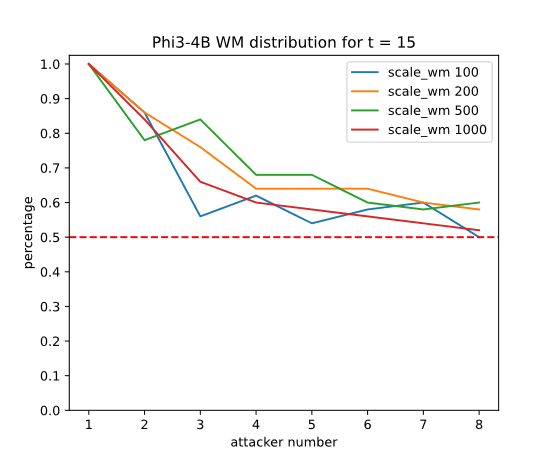}
        \end{minipage}
        \begin{minipage}{0.24\textwidth}
            \centering
            \includegraphics[width=\textwidth]{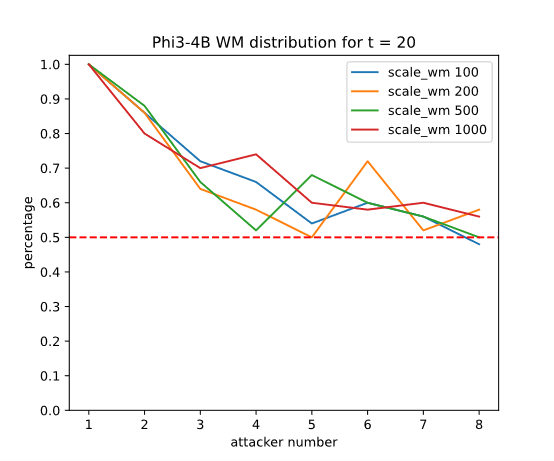}
        \end{minipage}
        
        \begin{minipage}{0.24\textwidth}
            \centering
            \includegraphics[width=\textwidth]{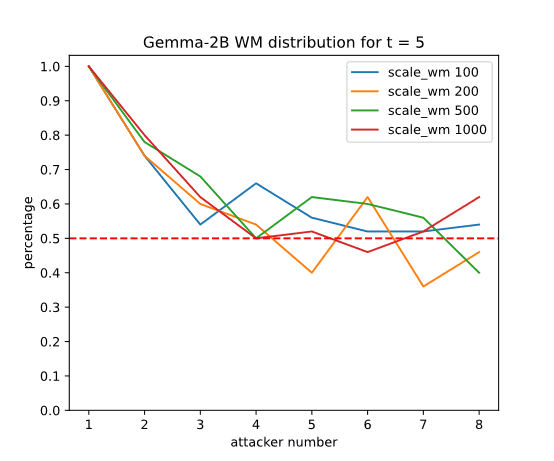}
        \end{minipage}
        \begin{minipage}{0.24\textwidth}
            \centering
            \includegraphics[width=\textwidth]{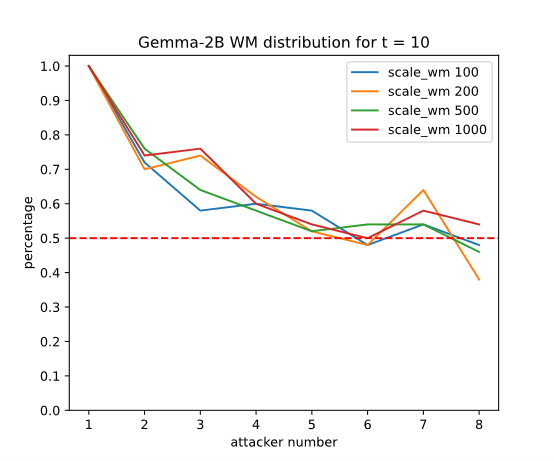}
        \end{minipage}
        \begin{minipage}{0.24\textwidth}
            \centering
            \includegraphics[width=\textwidth]{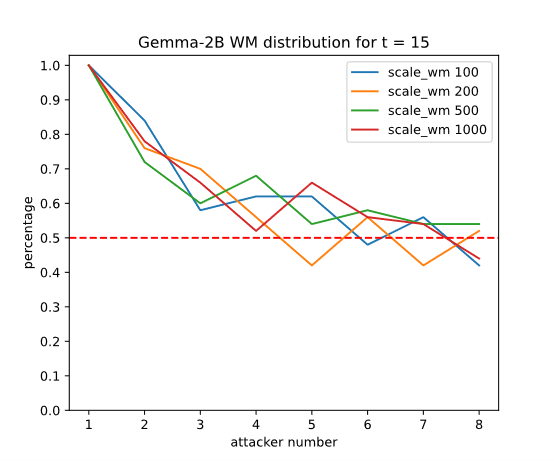}
        \end{minipage}
        \begin{minipage}{0.24\textwidth}
            \centering
            \includegraphics[width=\textwidth]{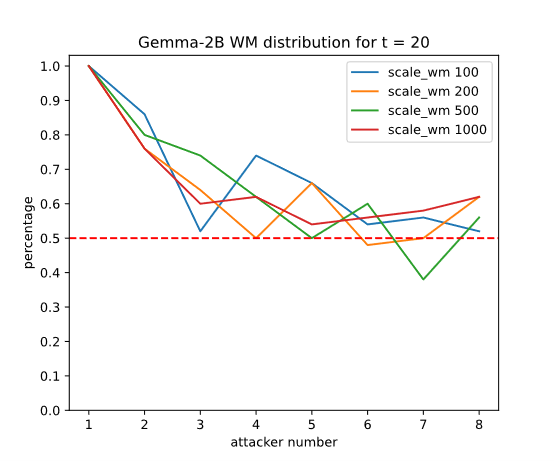}
        \end{minipage}
    \end{minipage}
    }
    \caption{Impact of colluding users on watermark extraction performance across varying lengths and scaling factors.}
    \label{fig:collusion}
\end{figure*}

\end{document}